\documentclass[epj]{svjour}
\usepackage{graphics}
\usepackage{hyperref}
\usepackage{epic,amsmath,graphicx}
\begin{document}
%
%\title{Radial Excitations of $D$, $D_s$, $B$, and $B_s$
%Heavy Mesons}
\title{Radial Excitations of Heavy-Light Mesons}
\author{Takayuki Matsuki\inst{1}, Toshiyuki Morii\inst{2},
\and Kazutaka Sudoh\inst{3}% etc
}                     % Do not remove
\institute{Tokyo Kasei University,
1-18-1 Kaga, Itabashi, Tokyo 173, JAPAN
\and Graduate School of Science and Technology,
Kobe University, Nada, Kobe 657-8501, JAPAN
\and Institute of Particle and Nuclear Studies, 
High Energy Accelerator Research Organization, \\ 
1-1 Ooho, Tsukuba, Ibaraki 305-0801, JAPAN}
%
%2006/10/23
\date{October 16, 2006}
\abstract{
Recent discovery of $D_s$ states suggests existence of radial excitations.
Our semirelativistic quark potential model succeeds in reproducing these states
within one to two percent of accuracy compared with the experiments,
$D_{s0}(2860)$ and $D_s^*(2715)$, which are identified as $0^+$ and $1^-$
radial excitations ($n=2$).
We also present calculations of radial excitations for $B/B_s$ heavy mesons.
Relation between our formulation and the modified Goldberger-Treiman relation is
also described.
\PACS{
      {12.39.Hg, 12.39.Pn, 12.40.Yx, 14.40.Lb, 14.40.Nd}
      {potential model; spectroscopy; heavy mesons}
%   \and
%      {PACS-key}{discribing text of that key}
     } % end of PACS codes
} %end of abstract
\maketitle
\section{Introduction}
\label{intro}
BaBar has recently announced the discovery of a new $D_s$ state, which seems to
be $c\bar s$ state,\cite{Palano06} $D_{s0}(2860)$.
Subsequent to this experiments Belle has observed a new state of $D_s^*(2715)$ whose
spin and parity is determined to be $1^-$.\cite{Abe06}

We were the first in predicting $0^+$ and $1^+$ states of $D_{s}$
and $D$ particles, $D_{s0}(2317)$, $D_{s1}(2460)$, $D_{0}^{*}(2308)$, and
$D_{1}'(2427)$,\cite{Matsuki97}, and we have also succeeded in reproducing higher
resonances of $B/B_s$
particles, $B_1(5720)$, $B_2^*(5745)$, and $B_{s2}^*(5839)$,\cite{Matsuki06} 
using our semirelativistic quark potential model. Our model succeeds in lowering
$0^+$ and $1^+$ states of $D_s$ so that these mass values are below $DK/D^*K$
threshold which other models do not easily succeed.
This model respects both heavy
quark symmetry and chiral symmetry in a certain limit of
parameters,\cite{Matsuki05} which relates our formulation with the idea of
the modified Goldberger-Treiman relation proposed in Refs.
\cite{Bardeen03,Harada04}.
Hence it is natural to try to explain the newly discovered $D_s$ states by using
our semirelativistic model and we are again successful in reproducing these
states discovered by BaBar and Belle.

To interpret the state $D_{s0}(2860)$, there are arguments that this $c\bar s$
state is explained to be a scalar by a coupled channel model\cite{Beveren06}, or
that it is a $J^P=3^-$ state\cite{Colangelo06}, or that it can be explained by
using a phenomenological interaction terms like our quark potential
model,\cite{Close06} or that it can be analyzed by using the $^3P_0$
model.\cite{Zhang06}

Starting from the astonishing discovery of $D_{sJ}$ particles with narrow decay
width by BaBar and CLEO, and confirmed by Belle, a series of successive
experiments on the spectrum of a heavy-light system, i.e., heavy mesons, heavy
quarkonium, and heavy baryons, stimulates theorists to explain all these spectra
as well as their decay modes. See the recent reviews of Refs.
\cite{Rosner06,Swanson06}.
It seems that a new era of spectroscopy is opening, which is challenging to
theorists to solve these spectra at the same time.

\section{Numerical Calculation}
\label{result}
Our model starts from a Hamiltonian with a scalar confining potential together with
a Coulombic vector potential, we expand the whole system, i.e., Hamiltonian, wave
function, and eigenvalue is expanded in $1/m_Q$, and we solve equations order by
order consistently. In the actual numerical calculation,
we expand a wave function in a power series of relative coordinate with
some weighting exponential times power functions.
A wave function has positive components of a heavy quark due to the lowest order
constraint when expanding in $1/m_Q$ and four components of a light antiquark.
Thus the lowest eigenfunction is two by four matrix. Other than angular part, its
radial part can be written like below.
\[
u_{k}(r), ~v_{k}(r)\sim w_{k}(r)\left(\frac{r}{a}\right)^{\lambda}
\exp \left[-(m_{q}+b)r-\frac{1}{2}\left(\frac{r}{a}\right)^{2} \right] ,
\]
where $k$ is a quantum number of an operator, 
$-\beta_{q} \left( \vec \Sigma_{q} \cdot \vec L + 1 \right)$
with $\vec \Sigma_q$ light quark spin and $\vec L$ light quark angular momentum,
which distinguishes uniquely each state\cite{Matsuki97},
for instance, $k=-1$ for a spin multiplet $(J^P=0^-, 1^-)$, $k=+1$ for $(0^+, 1^+)$,
etc., $u_{k}(r)$ and $v_{k}(r)$ are upper and lower components of radial wave
functions, and
%
%\[
$\lambda=\sqrt{k^{2} - \left(4\alpha_s/3\right)^{2}}$ ,
%\]
%
where $a$ and $b$ are included in a scalar potential as $S(r)=r/a^2+b$, $\alpha_s$ is
a strong coupling, and $w_{k}(r)$ is a finite series of a polynomial in $r$,
%
%\[
$w_{k}(r)=\sum_{i=0}^{N-1}a_{i}^{k}\left(r/a\right)^{i}$ ,
%\]
%
which takes different coefficients for $u_{k}(r)$ and $v_{k}(r)$. In actual
calculation, we have used $N=7$. Hence we can in principle obtain
seven different radial excitations.

In this paper, we calculate the most optimal values of parameters so that recently
discovered and known $n=2$ (the first radial excitation) $D_s$ particles are all
fitted well around
one percent of accuracy compared with the experiments. Here only a strong coupling
$\alpha_s$ is modified and other parameters are kept the same as in \cite{Matsuki06},
in which we have obtained $\alpha_s=0.261$ both for $D$ and $D_s$.
These are presented in Table \ref{parameter} at the first order of calculation in
$p/m_Q$ with $p$ being internal quark momentum and $m_Q$ heavy quark mass.

\begin{table}[t!]
\caption{Most optimal values of parameters.}
\label{parameter}
\begin{tabular}{lccc}
\hline
\hline
Parameters 
& ~~$\alpha_s^{n=2}$ & ~~$a$ (GeV$^{-1}$) & ~~$b$ (GeV) \\
%%%%%%%%%%%%%%%%%%%%%%%%%%%
%First order 
& ~~0.344 & ~~1.939 & ~~0.0749 \\
%%%%%%%%%%%%%%%%%%%%%%%%%%%
\hline
\hline
\end{tabular}
\end{table}
\begin{table}[t!]
%\caption{Most optimal values of parameters.}
%\label{parameter}
\begin{tabular}{cccc}
\hline
\hline
 ~~$m_{u, d}$ (GeV) & ~~$m_s$ (GeV) & ~~$m_c$ (GeV) & ~~$m_b$ (GeV) \\
%%%%%%%%%%%%%%%%%%%%%%%%%%%
%First order 
 ~~0.0112 & ~~0.0929 & ~~1.032 & ~~4.639 \\
%%%%%%%%%%%%%%%%%%%%%%%%%%%
\hline
\hline
\end{tabular}
\end{table}

With these values of parameters, we obtain $n=2$ masses of $D_s$ and $D$ at the
same time. The results are shown in Table \ref{Dsn2} for $D_s$. We also predict $n=2$
states for $D$, $B$, and $B_s$ states, which are shown in Tables \ref{Dn2}, \ref{Bn2},
and \ref{Bsn2} assuming the same strong coupling $\alpha_s$ for $D_s$, which may
actually be different for $B/B_s$ particles. In these Tables, $p_i$ and $n_i$
are $i-$th order positive and negative component contributions of a heavy quark,
respectively, and $c_i=p_i+n_i$.
When one carefully looks at these Tables, one notices that values of higher states
$^3D_1$ and $"^3D_2"$ are not reliable even though we have listed in Tables for
consistency with the former calculations.

\section{Mass Gap}
\label{mass_gap}
Noticing the mass gaps between spin multiplets, $(0^-, 1^-)$ and $(0^+, 1^+)$ for
$D_{sJ}$ mesons, are almost equal to each other, the modified Goldberger-Treiman
relation is proposed by Bardeen et al.\cite{Bardeen03,Harada04} to understand the
facts, i.e., the underlying physics might be chiral physics.
In other words the mass gap is caused due to the chiral symmetry breakdown which is
expressed by this relation. Further they have assumed hyperfine splittings due to
breakdown of heavy quark symmetry (inclusion of $1/m_Q$ corrections) are the
same within two spin multiplets, $(0^-, 1^-)$ and $(0^+, 1^+)$, so that the mass gap
is not affected by this hyperfine splitting.

In our formulation this is explained in Fig. \ref{mass_level}.\cite{Matsuki05}
Heavy quark symmetry reduces the original Hamiltonian into the one without spin
structure after projecting wave functions into positive and negative components of
a heavy quark.\cite{Matsuki97,Matsuki05}
Chiral symmetry of a light quark is realized by taking a limit of $m_q\rightarrow 0$
and $S(r)\rightarrow 0$, in
which case all the members of two spin multiplets, $(0^-, 1^-)$ and $(0^+, 1^+)$,
are degenerate. When the light quark mass
and a scalar potential are turned on, then the degeneracy due to chiral symmetry
is resolved and the mass gap corresponding to the modified Goldberger-Treiman
relation is given by
\[
  \Delta M = M_0(k=+1) - M_0(k=-1),
\]
where $M_0(k)$ is a degenerate mass for a quantum number $k$, which appears in Tables
\ref{Dsn2}$\sim$\ref{Bsn2} to distinguish states. This is described as
$\Delta M = \tilde g_\pi f_\pi/G_A$ in \cite{Bardeen03}. They have assumed dominant
interaction terms for hyperfine splitting due to $1/m_Q$ corrections so that mass gaps
between $0^+-0^-$ and $1^+-1^-$ are almost equal to each other, which seems
to hold in our formulation. Our hyperfine splittings due to $1/m_Q$ to this mass
gap $\Delta M$ are calculated and we have added those to degenerate mass gaps
between $0^+-0^-$ and $1^+-1^-$ for $D$, $D_s$, $B$, and $B_s$ with $n=1$, which are
given in Table \ref{massgap}. Our dynamical calculation supports the assumption that
the mass gaps between $0^+-0^-$ and $1^+-1^-$ are almost equal to each other in the
case of $n=1$. In the second row of the same Table values for $n=2$ are given, which
is not as good as the case for $n=1$.
\begin{table}[t!]
\caption{Theoretical mass gap. Values in brackets are experiments. Units are in MeV.}
\label{massgap}
\begin{tabular}{lcccc}
\hline
\hline
Mass gap ($n=1$)
& ~~$D$ & ~~$D_s$ & ~$B$ & ~$B_s$ \\
%%%%%%%%%%%%%%%%%%%%%%%%%%%
% gap between 0^+-0^-
$0^+-0^-$
& ~~414 (441) & ~~358 (348) & ~322 & ~239 \\
%%%%%%%%%%%%%%%%%%%%%%%%%%%
% gap between 0^+-0^-
$1^+-1^-$
& ~~410 (419) & ~~357 (348) & ~320 & ~242 \\
%%%%%%%%%%%%%%%%%%%%%%%%%%%
\hline
\hline
\end{tabular}
\end{table}
\begin{table}[t!]
\begin{tabular}{lcccc}
\hline
\hline
($n=2$)
& ~~$D$ & ~~$D_s$ & ~~$B$ & ~~$B_s$ \\
%%%%%%%%%%%%%%%%%%%%%%%%%%%
% gap between 0^+-0^-
$0^+-0^-$
& ~~308 & ~~274 & ~~206 & ~~160 \\
%%%%%%%%%%%%%%%%%%%%%%%%%%%
% gap between 0^+-0^-
$1^+-1^-$
& ~~350 & ~~327 & ~~216 & ~~171 \\
%%%%%%%%%%%%%%%%%%%%%%%%%%%
\hline
\hline
\end{tabular}
\end{table}

\vskip 10mm
\begin{table*}[h!]
\caption{$D_s (n=2)$ meson mass spectra (first order). Units are in MeV.}
\label{Dsn2}
\begin{tabular}{@{\hspace{0.5cm}}c@{\hspace{0.5cm}}c@{\hspace{1cm}}r@{\hspace{1cm}}r@{\hspace{1cm}}r@{\hspace{1cm}}c@{\hspace{1cm}}c@{\hspace{0.5cm}}}
\hline
\hline
$^{2s+1}L_J (J^P)$ & $M_0$ & 
\multicolumn{1}{c@{\hspace{1cm}}}{$c_1 /M_0$} & 
\multicolumn{1}{c@{\hspace{1cm}}}{$p_1 /M_0$} & 
\multicolumn{1}{c@{\hspace{1cm}}}{$n_1 /M_0$} & 
$M_{\rm calc}$ & $M_{\rm obs}$ \\
\hline
\multicolumn{1}{@{\hspace{0.6cm}}l}{$^1S_0 (0^-)$} 
& 2328 & 1.006 $\times 10^{-1}$ 
& 0.919 $\times 10^{-1}$ & 8.695 $\times 10^{-3}$ 
& 2563 & $-$ \\
\multicolumn{1}{@{\hspace{0.6cm}}l}{$^3S_1 (1^-)$} 
&  & 1.830 $\times 10^{-1}$ 
& 1.824 $\times 10^{-1}$ & 5.744 $\times 10^{-4}$ 
& 2755 & 2715 \\
\multicolumn{1}{@{\hspace{0.6cm}}l}{$^3P_0 (0^+)$} 
& 2456 & 1.553 $\times 10^{-1}$ 
& 1.470 $\times 10^{-1}$ & 8.245 $\times 10^{-3}$ 
& 2837 & 2856 \\
\multicolumn{1}{@{\hspace{0.6cm}}l}{$"^3P_1" (1^+)$} 
&  & 2.551 $\times 10^{-1}$ 
& 2.543 $\times 10^{-1}$ & 7.667 $\times 10^{-4}$ 
& 3082 & $-$ \\
\multicolumn{1}{@{\hspace{0.6cm}}l}{$"^1P_1" (1^+)$} 
& 2585 & 1.969 $\times 10^{-1}$ 
& 1.966 $\times 10^{-1}$ & 2.531 $\times 10^{-4}$ 
& 3094 & $-$ \\
\multicolumn{1}{@{\hspace{0.6cm}}l}{$^3P_2 (2^+)$} 
&  & 2.209 $\times 10^{-1}$ 
& 2.209 $\times 10^{-1}$ & 5.070 $\times 10^{-7}$ 
& 3157 & $-$ \\
\multicolumn{1}{@{\hspace{0.6cm}}l}{$^3D_1 (1^-)$} 
& 2391 & 8.605 $\times 10^{-1}$ 
& 8.600 $\times 10^{-1}$ & 5.016 $\times 10^{-4}$ 
& 4449 & $-$ \\
\multicolumn{1}{@{\hspace{0.6cm}}l}{$"^3D_2" (2^-)$} 
&  & -4.287 $\times 10^{-1}$ 
& -4.287 $\times 10^{-1}$ & 5.482 $\times 10^{-7}$ 
& 1366 & $-$ \\
%%%%%%%%%%%%%%%%%%%%%%%%%%%
\hline
\hline
\end{tabular}
\end{table*}
\vspace{-0.5cm}
\begin{table*}[h!]
\caption{$D (n=2)$ meson mass spectra (first order). Units are in MeV.}
\label{Dn2}
\begin{tabular}{@{\hspace{0.5cm}}c@{\hspace{0.5cm}}c@{\hspace{1cm}}r@{\hspace{1cm}}r@{\hspace{1cm}}r@{\hspace{1cm}}c@{\hspace{1cm}}c@{\hspace{0.5cm}}}
\hline
\hline
$^{2s+1}L_J (J^P)$ & $M_0$ & 
\multicolumn{1}{c@{\hspace{1cm}}}{$c_1 /M_0$} & 
\multicolumn{1}{c@{\hspace{1cm}}}{$p_1 /M_0$} & 
\multicolumn{1}{c@{\hspace{1cm}}}{$n_1 /M_0$} & 
$M_{\rm calc}$ & $M_{\rm obs}$ \\
\hline
\multicolumn{1}{@{\hspace{0.6cm}}l}{$^1S_0 (0^-)$} 
& 2241 & 1.078 $\times 10^{-1}$ 
& 0.975 $\times 10^{-1}$ & 1.038 $\times 10^{-2}$ 
& 2483 & $-$ \\
\multicolumn{1}{@{\hspace{0.6cm}}l}{$^3S_1 (1^-)$} 
&  & 1.917 $\times 10^{-1}$ 
& 1.910 $\times 10^{-1}$ & 6.882 $\times 10^{-4}$ 
& 2671 & $-$ \\
\multicolumn{1}{@{\hspace{0.6cm}}l}{$^3P_0 (0^+)$} 
& 2418 & 1.540 $\times 10^{-1}$ 
& 1.444 $\times 10^{-1}$ & 9.621 $\times 10^{-3}$ 
& 2791 & $-$ \\
\multicolumn{1}{@{\hspace{0.6cm}}l}{$"^3P_1" (1^+)$} 
&  & 2.493 $\times 10^{-1}$ 
& 2.488 $\times 10^{-1}$ & 5.352 $\times 10^{-4}$ 
& 3021 & $-$ \\
\multicolumn{1}{@{\hspace{0.6cm}}l}{$"^1P_1" (1^+)$} 
& 2491 & 2.076 $\times 10^{-1}$ 
& 2.070 $\times 10^{-1}$ & 5.956 $\times 10^{-4}$ 
& 3008 & $-$ \\
\multicolumn{1}{@{\hspace{0.6cm}}l}{$^3P_2 (2^+)$} 
&  & 2.319 $\times 10^{-1}$ 
& 2.318 $\times 10^{-1}$ & 1.101 $\times 10^{-4}$ 
& 3069 & $-$ \\
\multicolumn{1}{@{\hspace{0.6cm}}l}{$^3D_1 (1^-)$} 
& 2280 & 1.869 $\times 10^{-1}$ 
& 1.863 $\times 10^{-1}$ & 5.418 $\times 10^{-4}$ 
& 2706 & $-$ \\
\multicolumn{1}{@{\hspace{0.6cm}}l}{$"^3D_2" (2^-)$} 
&  & 2.100 $\times 10^{-1}$ 
& 2.099 $\times 10^{-1}$ & 1.203 $\times 10^{-4}$ 
& 2759 & $-$ \\
%%%%%%%%%%%%%%%%%%%%%%%%%%%
\hline
\hline
\end{tabular}
\end{table*}
\vspace{-0.5cm}
\begin{table*}[h!]
\caption{$B (n=2)$ meson mass spectra (first order). Units are in MeV.}
\label{Bn2}
\begin{tabular}{@{\hspace{0.5cm}}c@{\hspace{0.5cm}}c@{\hspace{1cm}}r@{\hspace{1cm}}r@{\hspace{1cm}}r@{\hspace{1cm}}c@{\hspace{1cm}}c@{\hspace{0.5cm}}}
\hline
\hline
$^{2s+1}L_J (J^P)$ & $M_0$ & 
\multicolumn{1}{c@{\hspace{1cm}}}{$c_1 /M_0$} & 
\multicolumn{1}{c@{\hspace{1cm}}}{$p_1 /M_0$} & 
\multicolumn{1}{c@{\hspace{1cm}}}{$n_1 /M_0$} & 
$M_{\rm calc}$ & $M_{\rm obs}$ \\
\hline
\multicolumn{1}{@{\hspace{0.6cm}}l}{$^1S_0 (0^-)$} 
& 5849 & 0.919 $\times 10^{-2}$ 
& 0.831 $\times 10^{-2}$ & 8.849 $\times 10^{-4}$ 
& 5902 & $-$ \\
\multicolumn{1}{@{\hspace{0.6cm}}l}{$^3S_1 (1^-)$} 
&  & 1.634 $\times 10^{-2}$ 
& 1.629 $\times 10^{-2}$ & 5.867 $\times 10^{-5}$ 
& 5944 & $-$ \\
\multicolumn{1}{@{\hspace{0.6cm}}l}{$^3P_0 (0^+)$} 
& 6025 & 1.375 $\times 10^{-2}$ 
& 1.289 $\times 10^{-2}$ & 8.590 $\times 10^{-4}$ 
& 6108 & $-$ \\
\multicolumn{1}{@{\hspace{0.6cm}}l}{$"^3P_1" (1^+)$} 
&  & 2.226 $\times 10^{-2}$ 
& 2.221 $\times 10^{-2}$ & 4.779 $\times 10^{-5}$ 
& 6160 & $-$ \\
\multicolumn{1}{@{\hspace{0.6cm}}l}{$"^1P_1" (1^+)$} 
& 6098 & 1.886 $\times 10^{-2}$ 
& 1.881 $\times 10^{-2}$ & 5.412 $\times 10^{-5}$ 
& 6213 & $-$ \\
\multicolumn{1}{@{\hspace{0.6cm}}l}{$^3P_2 (2^+)$} 
&  & 2.108 $\times 10^{-2}$ 
& 2.107 $\times 10^{-2}$ & 1.001 $\times 10^{-5}$ 
& 6227 & $-$ \\
\multicolumn{1}{@{\hspace{0.6cm}}l}{$^3D_1 (1^-)$} 
& 5888 & 1.610 $\times 10^{-2}$ 
& 1.605 $\times 10^{-2}$ & 4.668 $\times 10^{-5}$ 
& 5982 & $-$ \\
\multicolumn{1}{@{\hspace{0.6cm}}l}{$"^3D_2" (2^-)$} 
&  & 1.809 $\times 10^{-2}$ 
& 1.808 $\times 10^{-2}$ & 1.037 $\times 10^{-5}$ 
& 5994 & $-$ \\
%%%%%%%%%%%%%%%%%%%%%%%%%%%
\hline
\hline
\end{tabular}
\end{table*}
\vspace{-0.5cm}
\begin{table*}[h!]
\caption{$B_s (n=2)$ meson mass spectra (first order). Units are in MeV.}
\label{Bsn2}
\begin{tabular}{@{\hspace{0.5cm}}c@{\hspace{0.5cm}}c@{\hspace{1cm}}r@{\hspace{1cm}}r@{\hspace{1cm}}r@{\hspace{1cm}}c@{\hspace{1cm}}c@{\hspace{0.5cm}}}
\hline
\hline
$^{2s+1}L_J (J^P)$ & $M_0$ & 
\multicolumn{1}{c@{\hspace{1cm}}}{$c_1 /M_0$} & 
\multicolumn{1}{c@{\hspace{1cm}}}{$p_1 /M_0$} & 
\multicolumn{1}{c@{\hspace{1cm}}}{$n_1 /M_0$} & 
$M_{\rm calc}$ & $M_{\rm obs}$ \\
\hline
\multicolumn{1}{@{\hspace{0.6cm}}l}{$^1S_0 (0^-)$} 
& 5936 & 0.878 $\times 10^{-2}$ 
& 0.802 $\times 10^{-2}$ & 7.588 $\times 10^{-4}$ 
& 5988 & $-$ \\
\multicolumn{1}{@{\hspace{0.6cm}}l}{$^3S_1 (1^-)$} 
&  & 1.597 $\times 10^{-2}$ 
& 1.592 $\times 10^{-2}$ & 5.013 $\times 10^{-5}$ 
& 6031 & $-$ \\
\multicolumn{1}{@{\hspace{0.6cm}}l}{$^3P_0 (0^+)$} 
& 6063 & 1.399 $\times 10^{-2}$ 
& 1.325 $\times 10^{-2}$ & 7.429 $\times 10^{-4}$ 
& 6148 & $-$ \\
\multicolumn{1}{@{\hspace{0.6cm}}l}{$"^3P_1" (1^+)$} 
&  & 2.299 $\times 10^{-2}$ 
& 2.292 $\times 10^{-2}$ & 6.908 $\times 10^{-5}$ 
& 6202 & $-$ \\
\multicolumn{1}{@{\hspace{0.6cm}}l}{$"^1P_1" (1^+)$} 
& 6193 & 1.828 $\times 10^{-2}$ 
& 1.826 $\times 10^{-2}$ & 2.351 $\times 10^{-5}$ 
& 6306 & $-$ \\
\multicolumn{1}{@{\hspace{0.6cm}}l}{$^3P_2 (2^+)$} 
&  & 2.052 $\times 10^{-2}$ 
& 2.052 $\times 10^{-2}$ & 4.709 $\times 10^{-8}$ 
& 6320 & $-$ \\
\multicolumn{1}{@{\hspace{0.6cm}}l}{$^3D_1 (1^-)$} 
& 5999 & 7.631 $\times 10^{-2}$ 
& 7.627 $\times 10^{-2}$ & 4.449 $\times 10^{-5}$ 
& 6456 & $-$ \\
\multicolumn{1}{@{\hspace{0.6cm}}l}{$"^3D_2" (2^-)$} 
&  & -3.802 $\times 10^{-2}$ 
& -3.802 $\times 10^{-2}$ & 4.861 $\times 10^{-8}$ 
& 5770 & $-$ \\
%%%%%%%%%%%%%%%%%%%%%%%%%%%
\hline
\hline
\end{tabular}
\end{table*}
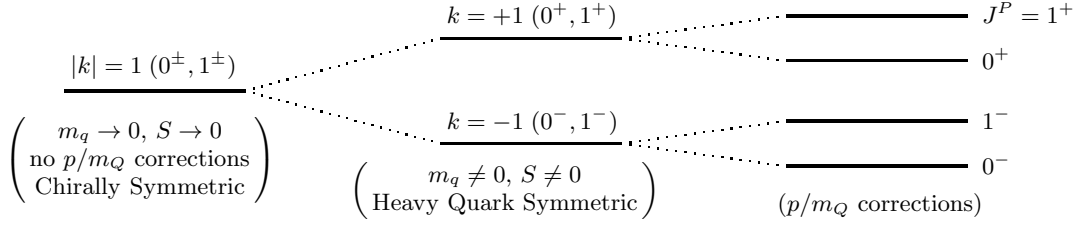
\begin{figure*}[t!]
\begin{center}
\begin{picture}(370,90)
\setlength{\unitlength}{0.4mm}
 \thicklines
  \put(5,40){\line(1,0){60}}
  \put(130,57.5){\line(1,0){60}} \put(130,22.5){\line(1,0){60}}
  \put(245,65){\line(1,0){60}} \put(245,50){\line(1,0){60}}
  \put(245,30){\line(1,0){60}} \put(245,15){\line(1,0){60}}
 \thinlines
  \dottedline{3}(65,40)(130,57.5) \dottedline{3}(65,40)(130,22.5)
  \dottedline{3}(190,57.5)(245,65) \dottedline{3}(190,57.5)(245,50)
  \dottedline{3}(190,22.5)(245,30) \dottedline{3}(190,22.5)(245,15)
  \put(7,46){$|k|=1\; (0^\pm, 1^\pm)$}
  \put(-15,15){$\left(\begin{array}{c}
              m_q\to0,\, S\to0 \\  {\rm no}~ p/m_Q~ {\rm corrections} \\
              {\rm Chirally~Symmetric}
             \end{array}\right)$}
  \put(132,62.5){$k=+1\;(0^+,1^+)$} \put(132,27.5){$k=-1\;(0^-,1^-)$}
  \put(310,62.5){$J^P=1^+$} \put(310,47.5){$0^+$}
  \put(310,27.5){$1^-$} \put(310,12.5){$0^-$}
%  \put(132,7.5){$\left(\begin{array}{c}
  \put(100,5){$\left(\begin{array}{c}
  			m_q\neq 0,\, S\neq0 \\
  			{\rm Heavy~Quark~Symmetric}
             \end{array}\right)$}
  \put(242,0){($p/m_Q$ corrections)}
\end{picture}
\caption{Procedure how the degeneracy is resolved in our model.
A quantum number $k$ is defined in the main text.}
\label{mass_level}
\end{center}
\end{figure*}

%%\begin{figure}
%%%\resizebox{0.75\textwidth}{!}{%
%%%  \includegraphics{leer.eps}
%\vspace{5cm}       % Give the correct figure height in cm
%%\caption{Please write your figure caption here}
%%\label{fig:1}       % Give a unique label
%%\end{figure}
%\begin{figure*}
%\vspace*{5cm}       % Give the correct figure height in cm
%\end{figure*}

%
% BibTeX users please use
% \bibliographystyle{}
% \bibliography{}

%%%%%%%%%%%%%%%%%% reference %%%%%%%%%%%%%%%%%%%
\def\Journal#1#2#3#4{{#1} {\bf #2}, #3 (#4)}
\def\NIM{Nucl. Instrum. Methods}
\def\NIMA{Nucl. Instrum. Methods A}
\def\NPB{Nucl. Phys. B}
\def\PLB{Phys. Lett. B}
\def\PRL{Phys. Rev. Lett.}
\def\PRD{Phys. Rev. D}
\def\PRO{Phys. Rev.}
\def\ZPC{Z. Phys. C}
\def\EPJ{Eur. Phys. J. C}
\def\PR{Phys. Rept.}
\def\IJM{Int. J. Mod. Phys. A}
\def\PTP{Prog. Theor. Phys.}

\end{document}